# Breaking through the bandwidth barrier in distributed fiber vibration sensing by sub-Nyquist randomized sampling

Jingdong Zhang, Tao Zhu*, Hua Zheng, Yang Kuang, Min Liu and Wei Huang

Key Laboratory of Optoelectronic Technology & Systems (Education Ministry of China) Chongqing University, Chongqing 400044, China

**ABSTRACT**

The round trip time of the light pulse limits the maximum detectable frequency response range of vibration in phase-sensitive optical time domain reflectometry (φ-OTDR). We propose a method to break the frequency response range restriction of φ-OTDR system by modulating the light pulse interval randomly which enables a random sampling for every vibration point in a long sensing fiber. This sub-Nyquist randomized sampling method is suits for detecting sparse-wideband-frequency vibration signals. Up to MHz resonance vibration signal with over dozens of frequency components and 1.153MHz single frequency vibration signal are clearly identified for a sensing range of 9.6km with 10kHz maximum sampling rate.

## 1. INTRODUCTION

Fiber optical distributed vibration sensing by phase-sensitive optical time domain reflectometry (φ-OTDR) has been intensively researched in numerous fields [1]. Practical sensing applications, involving leakage of pipeline, vibration of engine and creak of bridges or cranes, need long sensing range and high frequency response be satisfied simultaneously. For φ-OTDR systems, conventionally, the maximum frequency response range is determined by the round trip time of the probe light pulse, which is limited by the sensing fiber length. In order to overcome this problem, two kinds of approaches have been proposed recently. The first kind of schemes combines an interferometer and an φ-OTDR [2]. At the cost of weakening the multi-point detecting performance, the maximum frequency response of these systems are only limited by sampling frequency of data acquisition card (DAQ). Another approaches are based on frequency multiplexing which temporally sequenced multi-frequency probe light pulses are used to improve the sampling rate without decreasing the sensing range [3,4]. However, considering the sparse-wideband-frequency characteristic of vibration signal in many application fields, non-uniform randomized sampling shows aliasing proof in frequency domain and the capablility to identify multi-frequency signals with sub-Nyquist sampling rate [5,6].

In this letter, we propose a sub-Nyquist randomized sampling φ-OTDR by modulating the probe light pulse interval randomly. The random non-uniform sampling is analyzed and simulated thoroughly. A proof-of-concept random sampling φ-OTDR is performed and up to MHz multi-frequency vibration is identified at the equivalent sampling rate ranging from 5kHz to 10kHz.

## 2. PRINCIPLES AND SIMULATION

### 2.1 Uniform and non-uniform sampling

Over the years, theory and practice in the field of sampling have developed. According to different sampling intervals, there are two kinds of sampling methods, the uniform sampling and non-uniform sampling. The uniform sampling theorem, published by Shannon seven decades ago [7], requires that the sampling speeds be twice of the signal's maximal frequency component to avoid high-frequency aliasing. For a bandlimited signal $f(t)$ with maximal frequency $\omega_m$, the frequency spectrum $F_s(j\omega)$ of the uniform sampling signal with sampling rate $\omega_m$ is

$$F_s(j\omega) = \sum_{r=-\infty}^{\infty} F(j\omega - jr\omega_s) \tag{1}$$

*zhutao@cqu.edu.cn; phone +86 021 6511-1973; fax +86 021 6511-1973; oft.cqu.edu.cn

where $F(j\omega)$ is the frequency spectrum of the bandlimited signal $f(t)$. $\omega_s>2\omega_m$ should be at least guaranteed in the uniform sampling. This can be understood in time domain as shown in Fig. 1(a). The low frequency signal (thick black line) is overlapped with the high frequency signal (thin blue line) at the uniform sampling time. But if we swing the sampling interval slightly, these two signal are no longer superposition with each other. Actually, the non-uniform sampling is immune to frequency aliasing for this reason. As illustrated in Fig. 1(b), $dt$ is the random sampling interval with the maximum value $dt_{max}$ and the minimum value $dt_{min}$. For φ-OTDR systems, $dt_{min}$ is set to be the probe light pulse round trip time $t_r$ which equals to $2ln_{eff}/c$, where $l$ and $n_{eff}$ are the length and the effective index of sensing fiber, $c$ is the vacuum light speed. As shown in Fig. 1(c), $dt$ satisfies the (100us, 200us) uniform random distribution corresponding to a 10km length sensing fiber.

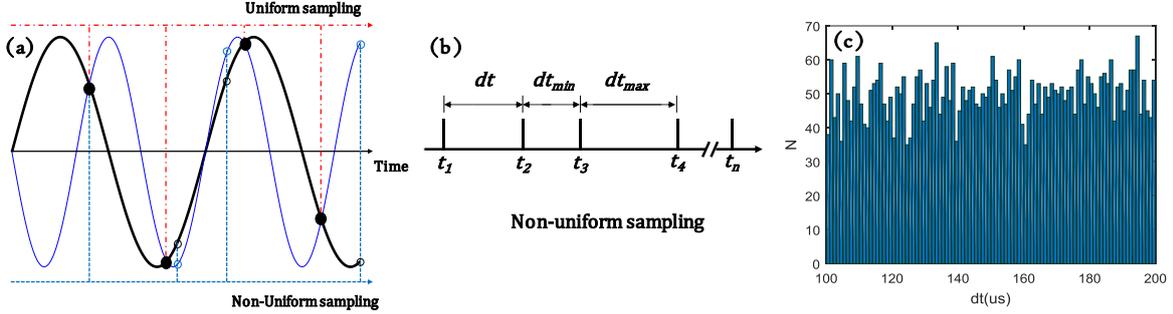

Figure 1. (a) Uniform and non-uniform sampling. Thick black line, low frequency signal. Thin blue line, high frequency signal. (b) Sampling time of random non-uniform sampling. $dt$, sampling interval. (c) Histogram bar chart of $dt$, $dt$ are uniform distributed random numbers.

Assumed the random sampling signal is $x(n)$ and $n=1, 2, 3…N$, the frequency properties can be obtained by taking non-uniform discrete Fourier transform (NDFT) of $x(n)$ [3]

$$X(\omega) = \sum_{n=1}^{N-1} x(n)\exp(-j2\pi\omega \cdot t_n)(t_{n+1} - t_n) \qquad (2)$$

## 2.2 Simulation

In order to test the performances of proposed method at sub-Nyquist rate, the uniform and random sampling are simulated and compared. The original signal is $x_0(t)=sin(2\pi f_1 t)+sin(2\pi f_2 t)+sin(2\pi f_3 t)$ where $f_1$=3kHz, $f_2$=61kHz and $f_3$=174kHz, as shown in Fig. 2(a). The 1000 points uniform sampling with 10 kHz sampling rate is performed and Fig. 2(b) shows the 10ms sampled data. Analyzed by taking fast Fourier transform (FFT), the frequency property of the sampled data is shown in Fig. 2(c) which the high frequency greater than 5 kHz is aliasing into 0 to 5kHz range. 1000 points random sampling with 10 kHz maximum sampling rate as illustrated in Fig. 1(c) is performed and Fig. 2(d) shows the 10ms sampled data. After taking NDFT from 0 to 200kHz, the three frequency components can be clearly identified in Fig. 2(e). It might be noted that the 1000 sampling points are used 50 times and the analysis bandwidth is 5kHz for each time. There is no frequency aliasing but background noise arises at full frequency span which can be suppressed by increasing the sampling length or adopting other modified non-uniform sampling method.

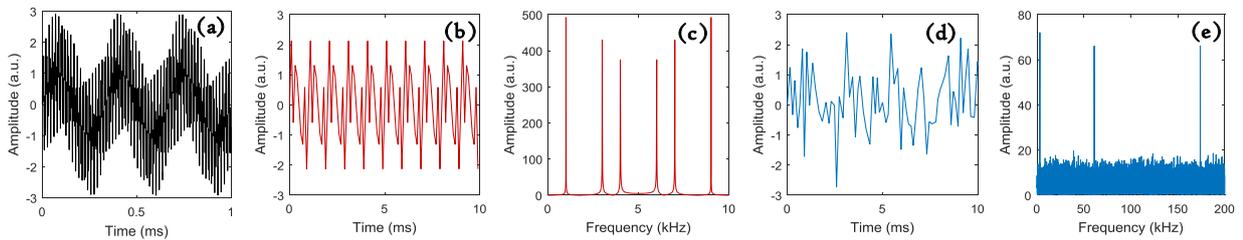

Figure 2. Mathematical simulation of uniform and sub-Nyquist random sampling (a) Original signal with 3kHz, 61kHz and 174kHz frequency component. (b) Uniform sampled data at 10kHz sampling rate. (c) Frequency spectrum form the 1000 points uniform sampled data. (d) Non-uniform sampled data. (e) Frequency spectrum from the 1000 points non-uniform sampled data.

After the mathematical simulation, a non-uniform data acquisition system is built to further evaluate the performances of the proposed method in practice. An arbitrary waveform generator (AWG, Tektronix, AWG5012C) is used to generate the random trigger signals which depicted in Fig. 1(b) and 1(c). Together with a data acquisition card (DAQ, Gage, CSE24G8), we test the data acquisition system by sampling a 20kHz to 190kHz sweep frequency signal step by step with 10kHz frequency interval. The number of sampling points is 1000 for every frequency step and the frequency spectrum is shown in Fig. 3(a).

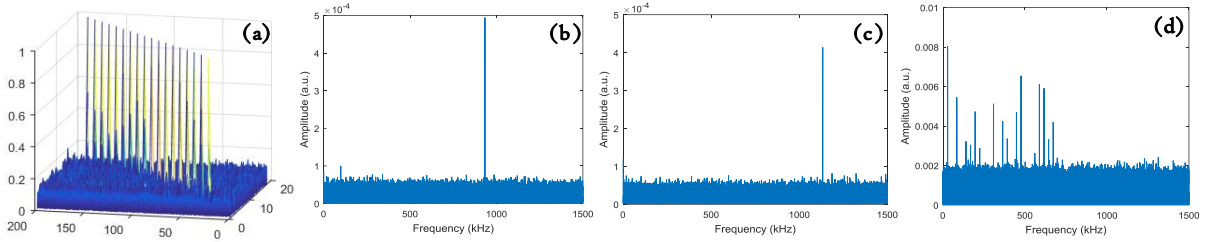

Figure 3. Detection of sweep frequency signal and vibration of PZT. (a) Frequency spectra of 20kHz to 190kHz sweep frequency signal with direct detection. (b-c) Frequency spectra of PZT vibration detected through MZI. The PZT is driven at 933kHz (b) and 1133kHz (c). (d) Frequency spectrum driven by the PZT around resonant frequency.

In order to simulate the vibration in the fiber sensing, a piezoelectric transducer (PZT) tube is adopted as the vibration source. The detection of PZT vibration is firstly performed in a Mach–Zehnder interferometer (MZI) where 2m length fiber is wrapped on the PZT tube and regarded as the sensing arm of the MZI. Typically, the respond of MZI to a sinusoidal vibration signal is $I(t) = \cos[VK\sin(\omega t) + \Delta\varphi]$ where $V$ is the driving voltage, $K$ is the response coefficient of PZT, $\omega$ is the vibration frequency and $\Delta\varphi$ is the phase different between the sensing arm and the reference arm. Driven the PZT at 933kHz and 1133kHz separately, 1000 points random samplings are record by the DAQ with the trigger of AWG. The corresponding frequency spectra are shown in Fig. 3(b-c) where the vibration frequency can be identified accurately. The PZT is resonant around 28kHz. When driven around the resonant frequency, the output intensity of MZI is a sparse-wideband-frequency signal with dozens of harmonic frequencies and can be detected by the proposed sub-Nyquist randomized sampling method. With driving the PZT at 28kHz, the frequency spectrum is shown in Fig 3(d), where over a dozen of harmonic frequency components can be observed.

## 3. EXPERIMENT AND RESULT

The experimental setup is shown in Fig. 4(a). A single frequency light source whose linewidth is less than 200Hz (NKT Laser, E15) is injected into an acoustic-optic modulator (AOM) driven by the AWG. The CW light from the laser is modulated into 50ns pulses with the pulses interval satisfying a (100us, 200us) uniform random distribution correspond to 10km sensing range. The modulated pulses are amplified through an Erbium-doped fiber amplifier (EDFA), the output of EDFA is split into two parts by using a 1:99 coupler. The 99 branch is injected into the 9.6km sensing fiber through the circulator which acts as the probe light pulse and the 1 branch is detected by a Photo Detector (PD) which provides the precise trigger source for the Data Acquisition card (DAQ, Gage, 2GSa/s). At the end of 9.6km sensing fiber, 2m length fiber is wrapped on the PZT tube. The Rayleigh backscattering light is amplified by anther EDFA and detected by a 350MHz low noise InGaAs PD (THORLABS, PDB430C), and recorded by the DAQ. Fig. 4(b-c) are the recorded Rayleigh backscattering signal of 9.6km sensing range and the end section of sensing fiber, separately.

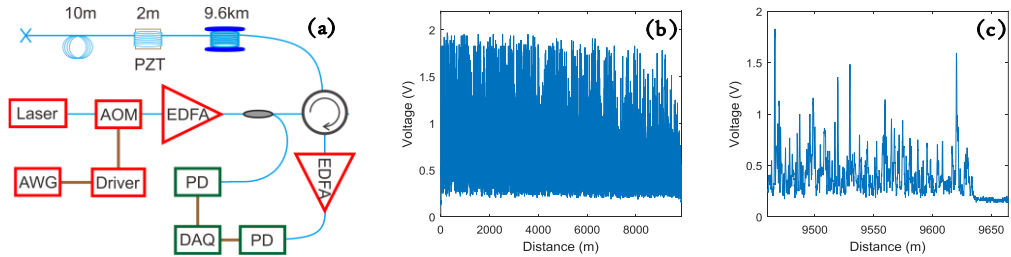

Figure 4. (a) Experimental setup. (b) The recorded Rayleigh backscattering signal of 9.6km sensing fiber. (c) The recorded Rayleigh backscattering signal at the end section of sensing fiber.

Driven the PZT at different frequency, the vibration frequency response of the proposed sub-Nyquist randomized sampling φ-OTDR system is analyzed by NDFT and the results are demonstrated below. Fig. 5(a) shows the first 30ms sampled signal of 4000 periods recorded signals at the end section of the sensing fiber while the PZT is driven at 450kHz. The vibration can be found ~ 9595m to 9600m and the spatial resolution is ~ 5m. Among the sampled vibration points, point at 9597.9m possesses the maximum stand deviation and the first 10ms sampled signal of this point is ploted in Fig. 5(b). Taking the 4000 points NDFT at 9597.9m, the frequency spectrum is shown in Fig. 5(c). Driven the PZT at 1.153MHz, the corresponding frequency spectra are indicated in Fig. 5 (d).

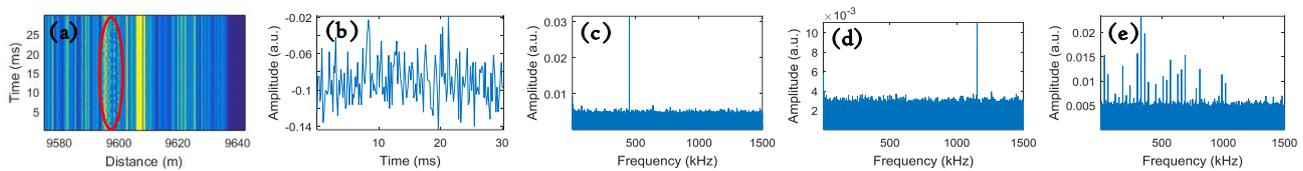

Figure 5. High frequency vibration detection results by proposed random sampling method (a). First 10ms signals of 4000 periods record signals at the end section when the PZT is driven at 450kHz. (b) First 10 ms sampled signal at 9597.9m. (c) and (e). Frequency spectra when driving the PZTs at 450 kHz and 1.153MHz. (e) Frequency spectrum when driving the PZT around resonant frequency.

When driving the PZT around resonant frequency, dozens of harmonic frequencies can be found at the frequency spectrum of vibration point, as shown in Fig. 5(e). This indicates that the proposed sub-Nyquist randomized sampling φ-OTDR system is able to detect the sparse-wideband high frequency vibration signals.

## 4. DISCUSSION

Thanks to the multiple advantageous natures, random sampling is a potential solution for high frequency detection in distributed fiber sensing with sub-Nyquist sampling rate. Since the redundancy is minimized compared with uniform sampling, random sampling is frequency aliasing proof and the sampled data of the same sampling cycle can be repeatedly used for frequency analyzing at arbitrarily frequency range. However, there is no fast algorithm for random sampling, the reconstruction of the original signal is difficult and the random sampling may bring noise background at full frequency span. These obstacles need be further studied if random sampling is practically applied in distributed fiber sensing.

## ACKNOWLEDGMENTS

This work is supported by the Project of Natural Science Foundation of China (Grant No: 61635004, 61475029, 61377066, 61405020), The science fund for distinguished young scholars of Chongqing (Grant No: CSTC2014JCYJJQ40002) and The Project of Natural Science Foundation of Chongqing (Grant No: cstc2013jcyjA40029).